\documentclass[onecolumn,aps,prb]{revtex4}
\usepackage{amsmath,amsfonts,bm,graphicx}

\newcommand{\s}{\sigma}

\newcommand{\ixe}{\rangle\!\langle}
\newcommand{\phd}{\phantom{\dagger}}

\begin{document}

\title{All-electric-spin control in
interference single electron transistors}

\author{Andrea Donarini}

\affiliation{Theoretische Physik, Universit\"{a}t Regensburg,
93040 Regensburg, Germany}

\author{Georg Begemann}

\affiliation{Theoretische Physik, Universit\"{a}t Regensburg,
93040 Regensburg, Germany}

\author{Milena Grifoni}

\affiliation{Theoretische Physik, Universit\"{a}t Regensburg,
93040 Regensburg, Germany}

\maketitle

\textbf{Single particle interference lies at the heart of quantum mechanics.
The archetypal double-slit experiment  \cite{Young1804} has been repeated with
electrons in vacuum \cite{Joensson61,Merli76} up to the more
massive $C_{60}$ molecules \cite{Arndt99}.
 Mesoscopic rings threaded by a magnetic flux provide the solid-state analogous
 \cite{Yacoby95, Gustavsson08}. Intra-molecular
interference  has been recently discussed in molecular junctions
\cite{Cardamone06, Ke08, Qian08, Begemann08,Darau09}. Here we propose
to exploit interference  to achieve all-electrical  control of a
single electron spin in quantum dots, a highly desirable property for
spintronics \cite{Wolf01,Awschalom07,Ohno00} and spin-qubit applications
\cite{Golovach06,Levitov03,Debald05,Walls07,Nowak07}.
The device consists of an interference single electron
transistor (ISET) \cite{Begemann08,Darau09}, where  destructive
interference between orbitally degenerate electronic states produces
 current blocking at specific bias voltages. We show that in the presence of
  parallel polarized ferromagnetic leads the interplay
between interference and the exchange coupling on the system
generates an effective energy renormalization yielding different blocking
biases for majority and minority spins. Hence, by tuning the bias
voltage full control over the spin of the trapped
electron is achieved.}

\begin{figure}
  \includegraphics[width=0.45\textwidth,angle=0]{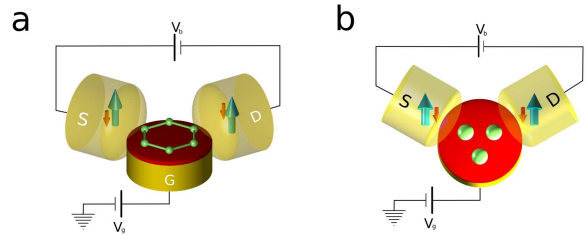}
  \caption{Two examples of interference single electron transistors (ISETs): a
benzene molecular junction contacted in the meta configuration (A) and a triple quantum
dot artificial molecule (B). The source and drain are parallel polarized ferromagnetic leads.}
  \label{fig1}
\end{figure}

The all-electrical solutions to the challenge of single spin control that have
been proposed \cite{Golovach06,Levitov03, Debald05,Walls07} and realized
\cite{Nowak07,Hauptmann08} are based either on spin orbit-coupling
\cite{Golovach06,Levitov03, Debald05,Walls07,Nowak07} or on tunneling-induced
spin splitting in the Kondo regime \cite{Hauptmann08}. Our proposal relies on
the current blocking occurring in an ISET due to interference between degenerate
states. The conditions for interference blocking are very generic \cite{Darau09}
and admit several different realizations. We consider here for clarity a benzene
and a
triple-dot ISET, Fig.~\ref{fig1}. Both are described by the Hamiltonian:
\begin{equation}
 H = H_{\rm sys} + H_{\rm leads} + H_{\rm T},
\end{equation}
where $H_{\rm sys}$ represents the central system and also contains the energy
shift operated by a capacitively coupled  gate electrode at the potential
$V_{\rm g}$. The Hamiltonian $H_{\rm sys}$ is in both cases invariant with
respect to a discrete set of rotations around the vertical axis passing through
the center of the system. This fact allows a classification of its eigenstates
in terms of the $z$ component of the angular momentum $\ell$ and also ensures
the existence of \emph{degenerate} states with different $\ell$. Then, a generic
eigenstate is represented by the ket $|N \ell \sigma E \rangle$ where $N$ is the
number of electrons on the system, $\sigma$ is the spin and $E$ the energy of
the state.  When degenerate states participate to transport they interfere
since, like the two paths of the double-slit experiment, they are occupied
simultaneously by the travelling electron, but in different superpositions under
diverse transport conditions. $H_{\rm lead}$ describes the ferromagnetic leads
with equal (for simplicity)  parallel polarization $P$ and with a difference
$eV_{\rm b}$ between their electrochemical potentials. Finally, $H_{\rm T}$
accounts for the weak tunnelling coupling between the system and the leads,
characteristic of SETs, and we consider the tunnelling events restricted to the
atoms or to the dots closest to the corresponding lead (Fig.~\ref{fig1}).
We explicitly
consider the Coulomb interaction only in the central part of the device (see the
supplementary material \ref{app:Hamiltonian}) due to the strong confinement
experienced there by the electrons while, apart from the polarization
assumption, we assume a non interacting approximation for the leads.

\begin{figure}
  \includegraphics[width=0.45\textwidth,angle=0]{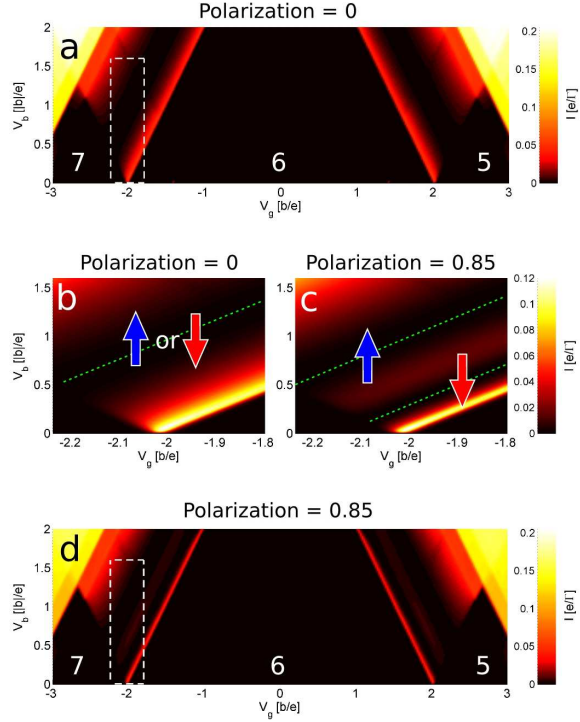}
  \caption{Benzene ISET: polarized vs. unpolarized configuration. Panel A -
Current vs. bias and gate voltage for unpolarized leads. Panel D - Current vs.
bias and gate voltage for polarized leads (polarization $P =  0.85$).  Panels B
and C - Blow up of the $6 \to 7$ particle transition for both configurations.
The unpolarized case shows a single current blocking line and the trapped
electron has either up or down polarization. The polarized case shows two
current blocking lines, corresponding to the different spin of the trapped
electron. The current is given in units of $e/\Gamma$ where $\Gamma$ is the bare
average rate (supplementary material \ref{app:frequencies}), and the temperature
$k_BT = 0.01b$ where $b$ is the hopping parameter (supplementary material
\ref{app:Hamiltonian})}
\label{fig2}
\end{figure}

\begin{figure}
  \includegraphics[width=0.45\textwidth,angle=0]{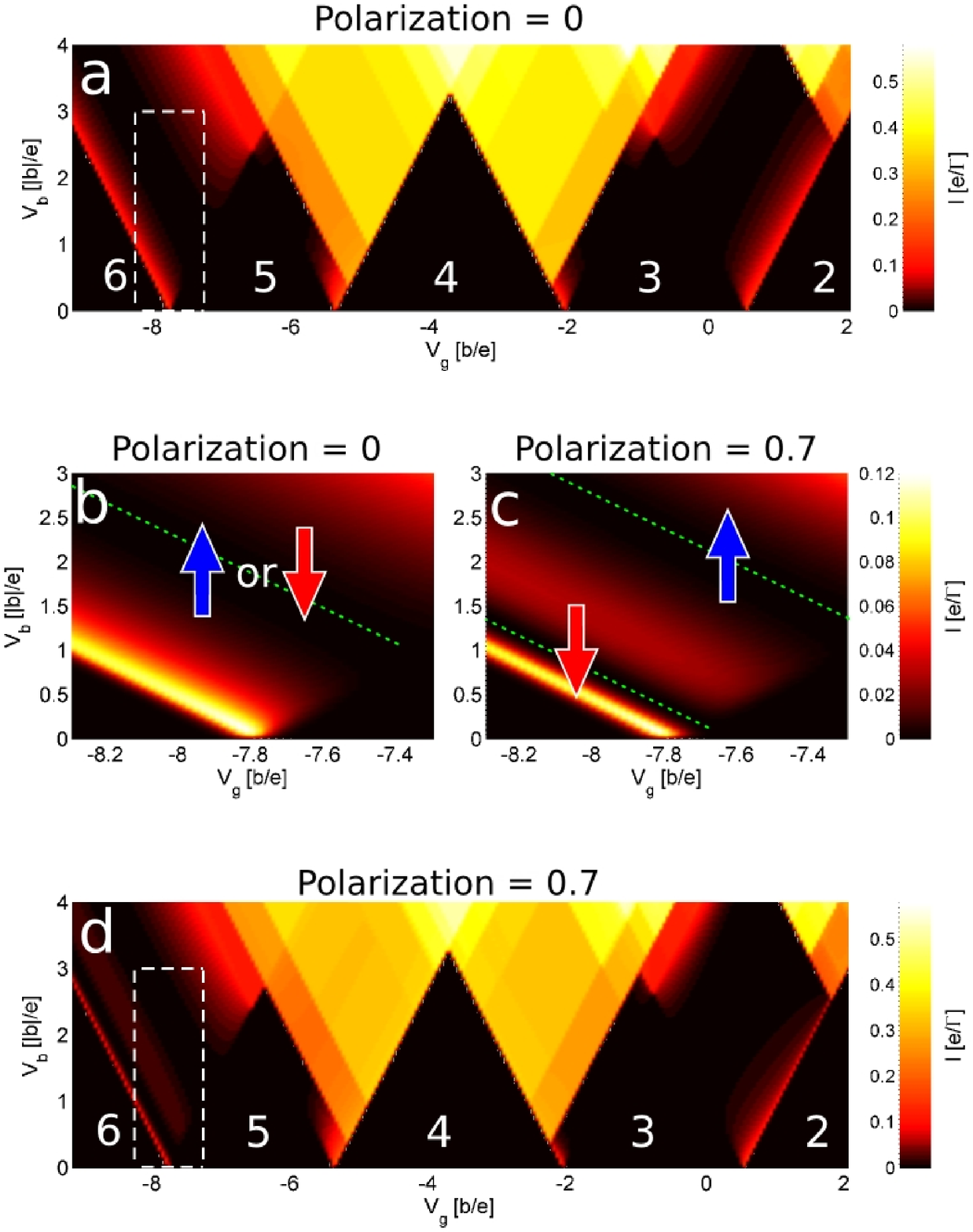}
  \caption{Triple dot ISET: polarized vs. unpolarized configuration. Panel A
- Current vs. bias and gate voltage for unpolarized leads. Panel D - Current
vs. bias and gate voltage for polarized leads (polarization $P =  0.7$). Panels
B and C -
Blow up of the $6 \to 5$ particle transition for both configurations.
The selective spin blocking is analogous to the one of the benzene ISET
(Fig.~\ref{fig2}).}
\label{fig3}
\end{figure}

In the weak coupling regime the current essentially
consists of sequential tunnelling events at the source and drain lead that
increase or decrease by one the number of electrons on the system.
The different panels of Figs.~\ref{fig2} and \ref{fig3} show the current
through the benzene and triple dot ISET, respectively, as a function of bias and
gate voltage. As in all SETs at low bias so called Coulomb diamonds, where
transport is energetically forbidden, occur. Within the diamonds the particle
number is fixed as indicated in the figures. Only exceptions are the charge
degenerate points where two diamonds meet. Here the energy difference of two
ground states with consecutive particle numbers is equal to the equilibrium
chemical potential of the leads.  At finite bias the incoming electrons have
enough energy to overcome the level spacing and the Coulomb repulsion and the
current flows. As a signature of the new states that enter the bias window, by
increasing the voltage the current typically increases steplike.

\begin{figure}[h!]
  \includegraphics[width=0.45\textwidth,angle=0]{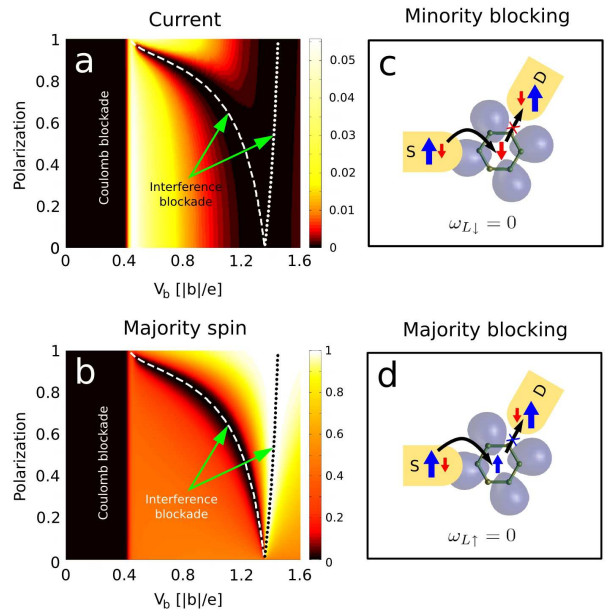}
  \caption{Spin control. Panel A - Current through the benzene ISET vs
bias and polarization at the $6 \to 7$ electrons transition. Panel B -
Population of the majority spin $7$ particle state. The two zero current lines
at high bias correspond to the maximum or minimum population of the $7$ particle
majority spin state and thus identify the spin state of the trapped electron on
the molecule. Panels C and D - Schematic representation of the spin selective
blocking corresponding to the dashed (C) and dotted (D) lines of the panels A
and B.}
  \label{fig4}
\end{figure}

In ISETs an exception to this picture is represented by the \emph{interference
blockade} where the current decreases for increasing bias generating negative
differential conductance (NDC) and eventually vanishes (see green lines in the
panels B and C of Fig.~\ref{fig2} and \ref{fig3}). Panels B in the same
figures indicate moreover that, for a given gate voltage and in absence of
polarization in the leads, the current is blocked only at  one specific bias
voltage. For parallel polarized leads, however, at a given gate voltage, the
current is blocked at \emph{two specific} bias voltages, one for each spin
configuration (panels C). As demonstrated below, the blocking of the minority
electrons occurs for the smaller bias voltages. As such full control of the spin
configuration in the ISET can be electrically achieved. The interference
blockade and its spin selectivity is also demonstrated in panels A and B of
Fig.~\ref{fig4}.  Along the dotted (dashed) line a majority (minority) spin
electron is trapped into the molecule. The molecular spin state can thus be
manipulated simply by adjusting the bias across the ISET. In the following we
discuss the physics of the spin-selective interference blocking and present the
necessary ingredients for its occurrence.

This novel blocking is explained by the presence of an $N$-particle
non-degenerate state and two degenerate $N+1$-particle states that
simultaneously contribute to transport. It also requires that the
ratio between the transition \emph{amplitudes} $\gamma_{\alpha i}$ ($i=1,2,\quad
\alpha=L,R$) between those $N$- and $N+1$-particle states is different for
tunneling at the left ($L$) and at the right ($R$) lead \cite{Darau09}:
\begin{equation}
\frac{\gamma_{L1}}{\gamma_{L2}} \neq \frac{\gamma_{R1}}{\gamma_{R2}}.
\label{eq:interf_condition}
\end{equation}
This condition is fulfilled in both cases presented in Fig.~\ref{fig1} due to
the geometrical configuration of the left and right lead. Due to condition
\eqref{eq:interf_condition}, the degenerate states interfere among themselves
such to form pairs of blocking and non-blocking states. The
blocking state is only coupled to the source lead (panels C and D of
Fig.~\ref{fig4}) while the non-blocking one to both source and drain. An
electron that populates the blocking state can neither leave towards the drain
nor, at high enough bias, return to the source since all energetically available
states are there filled. The non-blocking state is thus excluded from the
dynamics and the current vanishes.

As such we would conclude that the interference blocking is a threshold effect
and the current remains blocked until a new excited state participates to the
transport. However, as shown in Fig.~\ref{fig2} and \ref{fig3}, the current is
blocked only at specific values of the bias voltage. The explanation of this
phenomenon relies on two observations: i) The blocking state (Fig.~\ref{fig4})
must be antisymmetric with respect to the plane perpendicular to the system and
passing through its center and the atom (quantum dot) closest to the drain;
this state is thus also an eigenstate of the projection of the angular momentum
in the direction of the drain lead \cite{Eigenvalues}. At positive (negative)
bias voltages we call this state the $R(L)$-antisymmetric state
$|\psi_{R(L),\,a}\rangle$. ii) The coupling between the system and the leads not
only generates the tunneling dynamics described so far, but also contributes to
an internal dynamics of the system that leaves unchanged its particle number. In
fact the equation of motion for the reduced density matrix $\rho$ of the system
can be cast, to lowest non vanishing order in the coupling to the leads, in the
form:
\begin{equation}
 \dot{\rho} = -\frac{i}{\hbar} [H_{\rm sys},\rho] -
\frac{i}{\hbar}[H_{\rm eff},\rho] + \mathcal{L}_{\rm tun}\rho.
\label{eq:GME-small}
\end{equation}
The commutator with $H_{\rm sys}$ in Eq.~\eqref{eq:GME-small} represents
the coherent evolution of the system in absence of the leads. The operator
$\mathcal{L}_{\rm tun}$, describes instead the sequential tunnelling processes
and is defined in terms of the transition amplitudes $\gamma_{\alpha i}$ between
the $N$ and $N+1$ particle states like the ones introduced in equation
\eqref{eq:interf_condition}. Eventually  $H_{\rm eff}$ renormalizes the
coherent dynamics associated to the system Hamiltonian. It reads:
\begin{equation}
H_{\rm eff} = \sum_{\alpha\sigma} \omega_{\alpha\sigma} L_{\alpha},
\end{equation}
where $L_{\alpha}$ is the projection of the angular momentum in the
direction of the lead $\alpha$ and, for paramagnetic systems, it does not depend
on the spin degree of freedom $\sigma$. Moreover, $\omega_{\alpha\sigma}$ is the
frequency renormalization given to the states of spin $\sigma$ by their coupling
to the $\alpha$ lead. Equation \eqref{eq:GME-small} is an example of
Bloch-Redfield equation describing the dynamics of a system coupled to thermal
baths. A more detailed version of \eqref{eq:GME-small} is presented in the
supplementary material \ref{app:GME}.

For sake of simplicity we give in the following the explicit form of the
transition amplitudes $\gamma_{\alpha i}$, of the operator $L_{\alpha}$ and of
the associated frequency $\omega_{\alpha\sigma}$ only for the benzene ISET and
for the ground state transition $6_g \to 7_g$ that is characterized by
interference blocking. The argumentation is nevertheless very general and can be
repeated for all the systems exhibiting rotational symmetry. The transition
amplitudes read:

\begin{equation}
\gamma_{\alpha \ell} = \langle 6_g 0 0 | d_{M\sigma} | 7_g \ell \sigma
\rangle {\rm e}^{-i\ell\phi_\alpha},
\end{equation}
where $|7_g\,\ell\,\sigma \rangle$ are the orbitally degenerate $7$ particle
ground states,  $\ell = \pm 2$ the $z$ projection of the angular momentum in
units of $\hbar$ and $d^{\phd}_{{\rm M} \sigma}$ destroys an electron of spin
$\sigma$ in a reference  carbon atom $M$ placed in the middle between the two
contact atoms. Moreover, $\phi_\alpha$ is the angle of which we have to rotate
the molecule to bring the reference atom $M$ into the position of the
contact atom $\alpha$. The present choice of the reference atom implies that
$\phi_L = -\phi_R = \tfrac{\pi}{3}$. In the Hilbert space generated by the
two-fold orbitally degenerate $|7_g\,\ell\,\sigma\rangle$ the operator
$L_{\alpha}$ reads:

\begin{equation}
L_{\alpha}= \frac{\hbar}{2}
\left( \begin{array}{cc}
1 & e^{i2|\ell|\phi_{\alpha}}  \\
e^{-i2|\ell|\phi_{\alpha}} & 1 \\
 \end{array} \right).
\label{eq:angular_momentum}
\end{equation}

\begin{figure}[h!]
\vspace{0.5cm}
  \includegraphics[width=0.38\textwidth,angle=0]{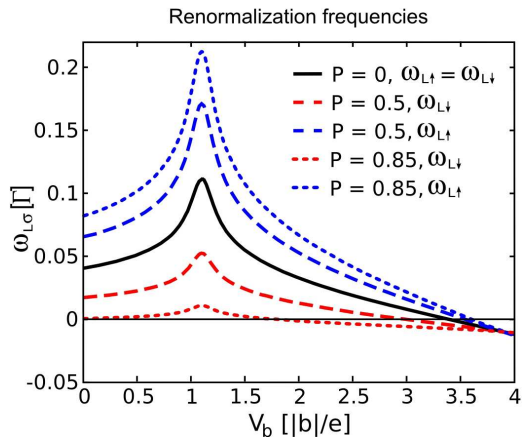}
  \caption{Blocking condition. Renormalization frequencies $\omega_{L \sigma}$
of a benzene ISET as
function of the bias and for different lead polarizations. The
current blocking condition $\omega_{L \sigma} = 0$ is fulfilled at different
biases
for the different spin states.}
  \label{fig5}
\end{figure}

For a derivation of (\ref{eq:angular_momentum}) see the supplementary material
\ref{app:L_alpha}
The frequency $\omega_{\alpha\sigma}$ is defined in terms of transition
amplitudes
to all the states of neighbour particle numbers:

\begin{equation}
\begin{split}
 \omega_{\alpha\sigma} =& \frac{1}{\pi}\sum_{\sigma' \{E\}}
\Gamma^0_{\alpha\sigma'}\\
 \Big[ & \langle 7_g\ell\sigma|d^{\phd}_{M \sigma'}|8 \{E\} \ixe 8\{E\}
|d^\dagger_{M \sigma'} | 7_g m \sigma \rangle p_\alpha(E - E_{7_g}) +\\
 & \langle 7_g\ell\sigma|d^\dagger_{M \sigma'}|6\{E\} \ixe 6\{E\}
|d^{\phd}_{M \sigma'} | 7_g m \sigma \rangle p_\alpha(E_{7_g} \!\!- E)\Big],
\end{split}
\label{eq:frequencies}
\end{equation}
where the compact notation $|N\{E\}\rangle$ indicates all possible states with
particle number $N$ and energy $E$, $p_\alpha(x)=-{\rm Re}\psi\left[\tfrac{1}{2}
+ \tfrac{i\beta}{2\pi}(x -\mu_\alpha)\right]$ where $\beta = 1/k_{\rm B}T$, $T$
is the temperature  and $\psi$ is the digamma function. Moreover
$\Gamma^0_{\alpha\sigma'} = \tfrac{2\pi}{\hbar}|t|^2 D_{\alpha\sigma'}$ is the
bare tunneling rate to the lead $\alpha$ of an electron of spin $\sigma'$, where
$t$ is the tunnelling amplitude and $D_{\alpha\sigma'}$ is density of states for
electrons of spin $\sigma'$ in the lead $\alpha$ at the corresponding chemical
potential $\mu_{\alpha}$. Due to the particular choice of the
arbitrary phase of the $7$ particle ground states, $\omega_{\alpha\sigma}$ does
not depend on the orbital quantum numbers $\ell$ and $m$. It depends instead on
the bias and gate voltage through the energy of the $6$, $7$-ground and $8$
particle states. In Fig.~\ref{fig5} the black curve depicts $\omega_{L\sigma}$
as
a function of the bias in absence of polarization: the frequencies corresponding
to the two spin species coincide and thus vanish at the same bias. The same
condition,

\begin{equation}
 \omega_{L\sigma} = 0,
\end{equation}
also determines the bias at which the current is completely blocked. In fact,
at that bias the effective Hamiltonian contains only the projection of the
angular momentum in the direction of the right lead (the drain) and the density
matrix corresponding to the full occupation of the $7$ particle
$R$-antisymmetric state ($\rho = |\psi_{R,\,a}\ixe \psi_{R,\,a}|$) is the
stationary
solution of Eq.~\eqref{eq:GME-small}. As we leave the blocking bias the
effective Hamiltonian contains also the projection of the angular momentum in
the direction of the left lead and the $R$-antisymmetric state is no longer
an
eigenstate of $H_{\rm eff}$. The corresponding density matrix is not a
stationary solution of \eqref{eq:GME-small} and current flows through the
system. The $L \leftrightarrow R$ symmetry of the system implies, for negative
biases, the blocking condition $\omega_{R\sigma} = 0$.

All-electric-spin control is achieved, in an ISET, only in presence of
ferromagnetic leads and with exchange interaction on the system. By manipulating
\eqref{eq:frequencies} it is possible to show that the frequency splitting
$\omega_{\alpha\uparrow} - \omega_{\alpha\downarrow}$ is proportional to the
polarization in the $\alpha$ lead, but vanishes in the absence of exchange
interaction on the system capable to lift the singlet-triplet degeneracy of the
excited $6$ and $8$ particle states (see the supplementary material
\ref{app:frequencies}). In Fig.~\ref{fig5} we show the frequencies
$\omega_{L\sigma} = 0$ vs. bias voltage also for a finite values of the
polarization $P$ calculated for the benzene ISET, where exchange splitting is
ensured by the strong Coulomb interaction on the system. The  interference blocking
conditions  $\omega_{L\sigma} = 0$ for the $L \to R$ current are satisfied at
different biases for the different spin species. The dotted and dashed lines in
Fig.~\ref{fig4} are the representation of the relations $\omega_{L\uparrow} =
0, \omega_{L\downarrow} = 0$ as a function of the bias and polarization,
respectively.

In previous studies \cite{Darau09} we have shown that the interference current
blocking does not depend on the perfect symmetry of the system but rather relies
on the existence of quasi-degenerate states in which the energy splitting is
smaller than the tunnelling coupling to the source and drain leads. In the
proposed structures the degeneracy is associated with the rotational symmetry
and it has the advantage of a simple geometrical realization of the
interference conditions \eqref{eq:interf_condition}. Nevertheless the effect is
more general and any other structure exhibiting orbital degeneracy is a good
candidate for an ISET.


We acknowledge financial support by the DFG under the programs SFB689, SPP1243.

\newpage
\appendix
\section*{Supplementary material}
\subsection{The system Hamiltonian}\label{app:Hamiltonian}

The Hamiltonian that describes both systems represented in Fig.~\ref{fig1}
reads
\begin{equation}
    \begin{split}
    H_{\rm sys} = \,\,
    &\xi_0 \sum_{i\s} d^{\dagger}_{i\s}d^{\phd}_{i\s} +
    b  \sum_{i\s}\left(d^{\dagger}_{i\s}d^{\phd}_{i+1\s} +
d^{\dagger}_{i+1\s}d^{\phd}_{i\s}\right)\\
    +& U \sum_i \left(n_{i\uparrow} - \tfrac{1}{2}\right)
                \left(n_{i\downarrow} - \tfrac{1}{2}\right)\\
    +& V \sum_i\left(n_{i\uparrow} + n_{i\downarrow}- 1\right)
               \left(n_{i+1\uparrow} + n_{i+1\downarrow} -
               1\right),
    \end{split}
    \label{eq:PPP}
\end{equation}
where $d^{\dagger}_{i\sigma}$ creates an electron of spin $\sigma$ in the $p_z$
orbital of carbon $i$ or in the ground state of the quantum dot $i$  and $i =
1,\ldots,\,6 (3)$ runs over the six carbon atoms (three quantum dots) of the
system. Moreover $n_{i\sigma} = d^{\dagger}_{i\sigma} d^{\phd}_{i\sigma}$. The
effect of the gate is included as a renormalization of the on-site energy $\xi =
\xi_0 - eV_{\rm g}$ ($V_{\rm g}$ is the gate voltage) and we conventionally set
$V_{\rm g} = 0$ at the charge neutrality point. The parameters that we have used
are $b = 2.5 eV,\, U= 9 eV,\, V= 6 eV.$

\subsection{The generalized master equation}\label{app:GME}
We describe the dynamics of the system with a generalized master
equation (GME) for the reduced density matrix $\rho$. This equation is obtained
from the Liouville equation for the full density matrix as a perturbation to the
lowest non vanishing order in the coupling to the leads by tracing out the leads
degrees of freedom. A generic formulation of the GME for a SET
in presence of degeneracies or quasi-degeneracies can be found
elsewhere {\it e.g.} \cite{Darau09}. We concentrate here on the range of gate
and bias voltages at which the dynamics is restricted to transitions involving
the $|6_g 0 0\rangle$  and $|7_g \ell \sigma\rangle$ many particle states of the
benzene ISET.

The seven particle states are spin and orbital degenerate. The general theory of
the GME would require  a priori to keep thus a full 4x4 density matrix
describing the 7 particle subspace. In presence of parallel polarized leads,
though, the coherences between different spin degrees of freedom can be
neglected since spin is always conserved by the electrons while travelling
through the device. The GME can thus be written in terms of the nine variables
collected in the 1x1 matrix $\rho^{6_g}$ and the two 2x2 matrices $\rho^{7_g
\sigma}$ with $\sigma =  \uparrow,\,\downarrow$. Due to the rotational symmetry
of the system it is more convenient to refer to another set of variables, namely
to describe the dynamics in terms of the occupation probabilities ${\rm W}_6$,
${\rm W}_{7\sigma}$ and the expectation values of the different projections of
the angular momentum for the system. The new set of variables is:
\begin{equation}
\begin{split}
{\rm W}_6 &= \rho^{6_g},\\
{\rm W}_{7 \sigma} &= {\rm Tr}\{\rho^{7_g\sigma}\},\\
{\rm L}_{\alpha\sigma} &= {\rm Tr}\{L_{\alpha}\rho^{7_g\sigma}\},\\
{\rm L}_{z\sigma} &={\rm Tr}\{L_z\rho^{7_g\sigma}\}.
\end{split}
\label{eq:variables}
\end{equation}
The operator $L_z$ is the generator of the set of discrete rotations around the
axis perpendicular to the plane of the benzene molecule that bring the molecule
into itself and can be written within the 7 particle Hilbert space spanned by
the vectors $|7_g\ell \sigma\rangle $ as $L_z = -\hbar |\ell|\sigma_z$, where
$\sigma_z$ is the third Pauli matrix. The operator $L_{\alpha}$ generates, in
the same space, the discrete rotations around the axis in the molecular plane
and passing through the center and the atom closest to the contact $\alpha$.
Finally, the dynamics for the variables introduced in
Eq.~\eqref{eq:variables} is given by the equations:
\begin{widetext}
\begin{equation}
\begin{split}
 \dot{\rm W}_6 =& 2\sum_{\alpha \sigma}\Gamma_{\alpha \sigma}
 \left[-f^+_\alpha(\Delta E){\rm W_6}
       +f^-_\alpha(\Delta E){\rm L}_{\alpha \sigma}\right],\\
 \dot{\rm W}_{7 \sigma} =& 2\sum_{\alpha}\Gamma_{\alpha \sigma}
 \left[f^+_\alpha(\Delta E)W_6
      -f^-_\alpha(\Delta E){\rm L}_{\alpha \sigma}\right],\\
\dot{\rm L}_{\alpha \sigma} =& -2\Gamma_{\alpha \sigma}f^-_{\alpha}(\Delta E)
 +2 \left\{\Gamma_{\alpha \sigma}f^+_{\alpha}(\Delta E)
         +\Gamma_{\bar{\alpha} \sigma}f^+_{\bar{\alpha}}(\Delta E)
\cos^2[|\ell|(\phi_\alpha - \phi_{\bar{\alpha}})]\right\}{\rm W}_6\\
         & + \Gamma_{\bar{\alpha} \sigma}f^-_{\bar{\alpha}}(\Delta E)
\sin^2[|\ell|(\phi_\alpha - \phi_{\bar{\alpha}})]{\rm W}_{7 \sigma}
           - \Gamma_{\bar{\alpha} \sigma}f^-_{\bar{\alpha}}(\Delta E)({\rm
L}_{\alpha \sigma} + {\rm L}_{\bar{\alpha} \sigma})
+\frac{\sin[2|\ell|(\phi_\alpha-\phi_{\bar{\alpha}})]}{4}\omega_{\bar{\alpha}
\sigma}{\rm
L}_{z \sigma},\\
\dot{\rm L}_{z \sigma} =& -\sum_{\alpha} \Gamma_{\alpha
\sigma}f^-_{\alpha}(\Delta E)
{\rm L}_{z \sigma}
- 2\tan[|\ell|(\phi_L - \phi_R)](\omega_{L \sigma} - \omega_{R \sigma})({\rm
W}_{7
\sigma} -
{\rm L}_{L \sigma} - {\rm L}_{R \sigma})\\
&- 2\cot[|\ell|(\phi_L - \phi_R)](\omega_{L \sigma} + \omega_{R \sigma})({\rm
L}_{L
\sigma} -
{\rm L}_{R \sigma}),
\end{split}
\end{equation}
\end{widetext}
where $\Gamma_{\alpha \sigma} = \Gamma^0_{\alpha \sigma}|\langle  6_g 0
0|d_{\alpha \sigma}|7_g \ell \sigma\rangle|^2$ is the tunnelling rate at the
lead $\alpha$ involving the ground states with 6 and 7 particles. Terms
describing sequential tunnelling from and to the lead $\alpha$ are proportional
to the Fermi functions $f^+_\alpha(x):=f(x - \mu_\alpha)$ and $f^-_\alpha(x) :=
1 -f^+_\alpha(x)$, respectively, and $\Delta E = E_{6g} - E_{7g} + eV_g$ where
$E_{6g}$ and $E_{7g}$ are the energies of the 6 and 7 particle ground states.
Finally with $\bar{ \alpha}$ we mean the lead opposite to the lead $\alpha$. By
using the expression $|\ell|$ (to be substituted with 2 for the $6 \to 7$
particle transition) we maintained the generality of the equations.  The
replacement $|\ell| = 2 \to  1$ and the appropriate redefinition of $\Delta E$
is enough to treat the $6 \to 5$ transition. Another important generalization
concerns the position of the leads. The para ($\phi_L - \phi_R = \pi$) and ortho
($\phi_L - \phi_R = \pi/3$) configuration are also treated within the same
equations. In particular one can see that all the terms containing the
renormalization frequencies drop from the equations in the para configuration
and that the equations for the ortho and meta configuration coincide.

\subsection{Matrix form of the operator $L_\alpha$}\label{app:L_alpha}

The explicit form of $L_{\alpha}$ is given in Eq.~\eqref{eq:angular_momentum}.
We give here its derivation. It is convenient, for this purpose, to choose the
arbitrary phases of the states $|7_g\ell \sigma\rangle$ in such a way that the
rotation of $\pi$ around the axis passing through a reference atom $M$ and the
center of the molecule transforms  $|7_g\ell \sigma\rangle$ into $-|7_g -\ell
\sigma\rangle$. In other terms
\begin{equation}
\exp(i\pi \tfrac{L_{M}}{\hbar}) = -\sigma_x,
\label{eq:rotation}
\end{equation}
where $\sigma_x$ is the first Pauli matrix. The relation is in fact an equation
for $L_M$ and the solution reads:
\begin{equation}
 L_M = \frac{\hbar}{2}
\left(
\begin{array}{cc}
1 & 1\\
1 & 1
\end{array}
\right).
\end{equation}
Eventually we obtain $L_\alpha$ by rotation of $L_M$ in the molecular plane,
namely:
\begin{equation}
L_{\alpha} =
e^{-\frac{i}{\hbar}\phi_\alpha L_z}
L_M
e^{\frac{i}{\hbar}\phi_\alpha L_z} = \frac{\hbar}{2}
\left( \begin{array}{cc}
1 & e^{i2|\ell|\phi_{\alpha}}  \\
e^{-i2|\ell|\phi_{\alpha}} & 1 \\
 \end{array} \right),
\end{equation}
where  $\phi_\alpha$ is the angle of which we have to rotate
the molecule to bring the reference atom $M$ into the position of the
contact atom $\alpha$.

\subsection{The spin splitting of the renormalization
frequencies}\label{app:frequencies}

The spin splitting of the renormalization frequencies is obtained from
Eq.~\eqref{eq:frequencies}. By introducing the average bare rate
$\Gamma = \tfrac{\Gamma^0_{\alpha\uparrow} +
\Gamma^0_{\alpha\downarrow}}{2}$, for simplicity equal in both
leads, and using the fact that benzene is paramagnetic we get:
\begin{equation}
\begin{split}
\omega_{\alpha\uparrow} -& \omega_{\alpha\downarrow} =
 2 \bar{\Gamma}^0_{\alpha} P_\alpha\frac{1}{\pi}\sum_{\{E\}}\\
 \Big[
 & \langle 7_g\ell\uparrow|d^{\phd}_{M \uparrow}|8 \{E\} \ixe 8\{E\}
|d^\dagger_{M \uparrow} | 7_g m \uparrow \rangle p_\alpha(E - E_{7_g})\\
 +& \langle 7_g\ell\uparrow|d^\dagger_{M \uparrow}|6\{E\} \ixe 6\{E\}
|d^{\phd}_{M \uparrow} | 7_g m \uparrow \rangle p_\alpha(E_{7_g} \!\!- E)\\
 -& \langle 7_g\ell\uparrow|d^{\phd}_{M \downarrow}|8 \{E\} \ixe 8\{E\}
|d^\dagger_{M \downarrow} | 7_g m \uparrow \rangle p_\alpha(E - E_{7_g}) \\
 -& \langle 7_g\ell\uparrow|d^\dagger_{M \downarrow}|6\{E\} \ixe 6\{E\}
|d^{\phd}_{M \downarrow} | 7_g m \uparrow \rangle p_\alpha(E_{7_g} \!\!- E)
\Big],
\end{split}
\end{equation}
where one appreciates the linear dependence of the spin splitting on the lead
polarization $P_\alpha$.
The first and the third term of the sum would cancel each other if the energy of
the singlet and triplet 8 particle states would coincide. An analogous
condition, but this time on the 6 particle states, concerns the second and the
fourth terms. For this reason the exchange interaction on the
system is a necessary condition to obtain spin splitting of the renormalization
frequencies and thus the full all-electric spin control.

\end{document}